\documentclass[apjl,twocolappendix]{emulateapj}
\usepackage[breaklinks,colorlinks,citecolor=blue]{hyperref}
\usepackage[all]{hypcap}
\usepackage[usenames,dvipsnames]{xcolor}
\usepackage{verbatim,epstopdf,graphicx,bigints, amsmath, tabularx}
\usepackage{array,afterpage,threeparttable}
\newcolumntype{L}[1]{>{\raggedright\hspace{0pt}}m{#1}}
\newcolumntype{C}[1]{>{\centering\hspace{0pt}}m{#1}}

\newcommand{\myemail}{bsalmon@stsci.edu}

\newcommand{\Mstar}{\hbox{$\mathrm{M}_\star$}}

\newcommand{\hst}{\textit{HST}}
\newcommand{\jwst}{\textit{JWST}}
\newcommand{\spitzer}{\textit{Spitzer}}

\usepackage{courier}
\def\code#1{\texttt{#1}}

\def\Snospace~{\S{}}

\hypersetup{
    colorlinks=true,
    linkcolor=blue,
    filecolor=magenta,      
    urlcolor=cyan,
}
\definecolor{venetianred}{rgb}{0.78, 0.03, 0.08}

\shorttitle{A Candidate $z\sim 10$ Arc}
\shortauthors{Salmon et al.}

\begin{document}
\title{A Candidate $z\sim10$ Galaxy Strongly Lensed into a Spatially Resolved Arc}
\author{Brett Salmon$^{1,\dagger}$, 
Dan Coe$^{1}$, 
Larry Bradley$^{1}$, 
Marusa Brada{\v c}$^{2}$,
Kuang-Han Huang$^{2}$,
Victoria Strait$^{2}$,
Pascal Oesch$^{3}$,
Rachel Paterno-Mahler$^{4}$,
Adi Zitrin$^{5}$,
Ana Acebron$^{5}$,
Nath\'alia Cibirka$^{5}$,
Shotaro Kikuchihara$^{6,7}$,
Masamune Oguri$^{7,8,9}$,
Gabriel B. Brammer$^{1}$,
Keren Sharon$^{4}$,
Michele Trenti$^{10}$,
Roberto J. Avila$^{1}$,
Sara Ogaz$^{1}$,
Felipe Andrade-Santos$^{11}$,
Daniela Carrasco$^{10}$,
Catherine Cerny$^{4}$,
William Dawson$^{12}$,
Brenda L. Frye$^{13}$,
Austin Hoag$^{2}$,
Christine Jones$^{11}$,
Ramesh Mainali$^{13}$,
Masami Ouchi$^{6,8}$,
Steven A. Rodney$^{14}$,
Daniel Stark$^{13}$,
Keiichi Umetsu$^{15}$
}
\affil{
$^{1}$Space Telescope Science Institute, Baltimore, MD, USA \\
$^{2}$Department of Physics, University of California, Davis, CA 95616, USA \\
$^{3}$Geneva Observatory, University of Geneva, Ch. des Maillettes 51, 1290 Versoix, Switzerland \\
$^{4}$Department of Astronomy, University of Michigan, 1085 South University Ave, Ann Arbor, MI 48109, USA \\
$^{5}$Physics Department, Ben-Gurion University of the Negev, P.O. Box 653, Beer-Sheva 84105, Israel \\
$^{6}$Institute for Cosmic Ray Research, The University of Tokyo, 5-1-5 Kashiwanoha, Kashiwa, Chiba 277-8582, Japan \\
$^{7}$Department of Physics, Graduate School of Science, The University of Tokyo, 7-3-1 Hongo, Bunkyo-ku, Tokyo 113-0033, Japan \\
$^{8}$Kavli Institute for the Physics and Mathematics of the Universe (Kavli IPMU, WPI), University of Tokyo, Chiba 277-8582, Japan \\
$^{9}$Research Center for the Early Universe, The University of Tokyo, 7-3-1 Hongo, Bunkyo-ku, Tokyo 113-0033, Japan \\
$^{10}$School of Physics, University of Melbourne, VIC 3010, Australia \\
$^{11}$Harvard-Smithsonian Center for Astrophysics, 60 Garden Street, Cambridge, MA 02138, USA \\
$^{12}$Lawrence Livermore National Laboratory, P.O. Box 808 L- 210, Livermore, CA, 94551, USA \\
$^{13}$Department of Astronomy, Steward Observatory, University of Arizona, 933 North Cherry Avenue, Rm N204, Tucson, AZ, 85721, USA \\
$^{14}$Department of Physics and Astronomy, University of South Carolina, 712 Main St., Columbia, SC 29208, USA \\
$^{15}$Institute of Astronomy and Astrophysics, Academia Sinica, PO Box 23-141, Taipei 10617, Taiwan \\
}
\altaffiltext{}{$\dagger\ $\myemail}
\submitted{Submitted to ApJL on 1/9/17}

\begin{abstract} 
The most distant galaxies known are at $z\sim10-11$, observed $400-500$ Myr 
after the Big Bang. The few $z\sim10-11$ candidates discovered to date have 
been exceptionally small-- barely resolved, if at all, by the \emph{Hubble~Space~Telescope}.
Here we present the discovery of SPT0615-JD, a fortuitous $z\sim10$ 
($z_{\rm phot}$=$9.9\pm0.6$) galaxy candidate stretched into an arc over $\sim2.5\arcsec$  
by the effects of strong gravitational lensing. 
Discovered in the Reionization Lensing Cluster Survey (RELICS) 
\emph{Hubble} Treasury program and companion S-RELICS \emph{Spitzer} program,
this candidate has a lensed $H$-band magnitude of $25.7\pm0.1$~AB~mag. 
With a magnification of $\mu\sim4-7$ estimated from our lens models, the
de-lensed intrinsic magnitude is $27.6\pm0.3$~AB~mag, and the half-light radius is 
$r_e<0.8$~kpc, both consistent with other $z>9$ candidates.
The inferred stellar mass ($\log [M_\star /\rm{M}_\Sun]=9.7^{+0.7}_{-0.5}$) and star formation rate 
($\log [\rm{SFR}/{\rm{M}}_\Sun$~${\rm{yr}}^{-1}]=1.3^{+0.2}_{-0.3}$) indicate that this candidate is a typical 
star-forming galaxy on the $z>6$ SFR--$M_\star$ relation. 
We note that three independent lens models predict two counterimages, 
at least one of which should be of a similar magnitude to the arc, but
these counterimages are not yet detected. Counterimages would not be expected if the arc 
were at lower redshift. However, the only spectral energy distributions capable of fitting 
the \emph{Hubble} and \emph{Spitzer} photometry well at lower redshifts require unphysical 
combinations of $z\sim2$ galaxy properties. The unprecedented lensed size of this 
$z\sim10$ candidate offers the potential for the \emph{James~Webb~Space~Telescope} to 
study the geometric and kinematic properties of a galaxy observed 500~Myr after the Big Bang. 
\end{abstract}

\section{Introduction} 
With its high resolution and sensitivity, observations using the 
\emph{Hubble Space Telscope} (\hst) have sharpened our understanding 
of the high-$z$ universe. Deep and wide
extragalactic imaging surveys with ACS and WFC3 have uncovered
thousands of galaxies at $z > 6$ in blank fields 
\citep[see][for reviews]{Finkelstein16,Stark16},
including the most distant galaxy found to-date at $z=11.1$ \citep[GN-z11;][]{Oesch16}. In
addition, we have prioritized \hst\ to observe the most massive galaxy
clusters, taking advantage of the natural telescopes they create via
strong gravitational lensing (CLASH, PI Postman; Frontier Fields, PI
Lotz; RELICS, PI Coe). This investment in lensing fields has proven 
fruitful. We have discovered highly magnified 
(MACS1149-JD, \citealt{Zheng12}, \citealt{Hoag17};  
MACS1115-JD and MACS1720-JD, 
\citealt{Bouwens14}; MACS0416-JD, \citealt{Infante15})
and multiply-imaged galaxies
(MACS0647-JD, \citealt{Coe13}; A2744-JD, \citealt{Zitrin14}) at redshifts up to $z \sim10.8$,
which have allowed us to study faint UV metal lines \citep{Stark14,Rigby15,Mainali17}, 
nebular emission lines \citep{Smit17a, Stark15b, Hoag17, Laporte17}, and
the star formation rate density deep into the epoch of reionization 
\citep{Oesch14,Oesch17}. 

However, little is known in detail about the $z>9$ universe, and the
handful of candidates found so far exhibit peculiar properties.
At $z\sim11$, MACS0647-JD has a radius smaller than 100 pc, the size of Giant
Molecular Clouds in the local universe. GN-z11 is three times brighter
than the characteristic UV luminosity ($L_*$) of galaxies at that
distance, surprisingly bright given the CANDELS search area. Both z$\sim$10
candidates MACS1149-JD and M0416-JD appear to have an evolved stellar
population of $\approx 340$ Myr (due to red [3.6~\micron ]$-$[4.5~\micron ] \spitzer\ colors), 
when the age of the universe was only
$\approx500$ Myr \citep{Hoag17}. \emph{JWST} NIRCam will better sample
the rest-frame UV-to-optical colors which will break some parameter 
degeneracies and challenge these initial inferences. However, with typical 
$z\sim10$ effective radii of $<0.2$\arcsec\ and a NIRCAM PSF FWHM\footnote{see
\url{https://jwst-docs.stsci.edu}} of 
$\sim0.05$\arcsec\ at 1.5~\micron, it will still be difficult resolve these galaxies spatially. 
Ideally, we can use the help of strong lensing to study the kinematics 
and intrinsic stellar populations at $z \sim10$ in detail. 

In this Letter we present a galaxy gravitationally lensed into an arc
with a photometric redshift of $z_{\rm{phot}}=9.9\pm0.6$.
Discovered in the Reionization Lensing Cluster Survey (RELICS) 
\emph{Hubble} (\hst) and \emph{Spitzer Space Telescope} imaging, the arc features of
this candidate extend across $\sim$2.5\arcsec , allowing unprecedented 
physical resolution deep in the epoch of reionization. 
This new candidate has an \hst\ F160W 
$H$-band magnitude of $H$=25.7$\pm0.1$ AB, bright enough for follow-up
spectroscopic or grism observations. In this work, we present 
the supporting evidence that this candidate is indeed at $z\sim10$, and discuss 
the remaining uncertainties. Throughout, we assume concordance cosmology
with $H_0$=70 km s$^{-1}$~Mpc$^{-1}$, $\Omega_{\Lambda,0}$=0.7 
and $\Omega_{\rm M,0}$=0.3.

\section{Data and Photometry}\label{sec:Data}
The galaxy cluster SPT-CL J0615-5746 (hereafter SPT0615-57; also known as PLCK 
G266.6-27.3) was discovered independently by the South Pole Telescope survey 
\citep{Williamson11} and the \cite{Planck11}. 
It is exceptionally massive ($M_{500} = 6.8\times10^{12}\ M_\Sun$) for its high redshift 
($z=0.972$). The SPT and Planck teams obtained \hst\ imaging (GO 12477 and 12757)
of the cluster with the ACS/WFC F606W filter $V$ (1-orbit depth) and F814W filter $I$ 
(combined 2-orbit depth). RELICS (GO 14096) obtained ACS/WFC imaging (1 orbit) 
in F435W $B$ and WFC3/IR imaging
(2 orbits) in F105W $Y$, F125W $J$, F140W $JH$, and F160W $H$. 

RELICS obtained similar \hst\ imaging with WFC3/IR and ACS as needed on 
a total of 41 clusters. The details of the image reduction,
SExtractor \citep[version 2.8.6;][]{Bertin96} object selection, and \hst\ photometry 
are described by \cite{Salmon17} and Coe et al. (in prep). SPT0615-57 was the second highest
high-$z$-producing cluster field out of the 41 RELICS fields, revealing 25 new candidate galaxies 
over the redshift range ${5.5<z<8.5}$ \citep{Salmon17}.

\bgroup
\def\arraystretch{1.3}%
\begin{table}
\caption{RELICS $z\sim$ 10 Candidate and $z\sim3$ Interlopers}\label{tab:candidates}
\smallskip
\begin{tabular*}{\columnwidth}{l @{\extracolsep{\fill}}l l l}
\toprule \\ [-0.2cm]
Field                                          &  SPT0615-57         & PLCKG138-10          & RXC0018+16           \\
RELICS ID                                      &  336                & 748                  & 1107                 \\
$\alpha_{J2000}$                               &  06:15:55.03        & 02:27:00.86          & 00:18:33.84          \\
$\delta_{J2000}$                               &  $-$57:46:19.56     & 49:00:22.68          & 16:25:18.84          \\
$B_{435}$                                      &  $>$28.7            & $>$26.8              & $>$28.8              \\
$V_{606}$                                      &  $>$28.4            & $>$28.4              & $>$28.8              \\ 
$I_{814}$                                      &  $>$29.5            & $>$27.0              & $>$29.4              \\ 
$Y_{105}$                                      &  $>$27.3            & $>$27.3              & $>$28.4              \\
$J_{125}$                                      &  $>$26.5            & $>$26.5              & $>$26.9              \\  
$JH_{140}$                                     &  $>$26.3            & 26.0$\pm$0.2         & $>$26.6              \\     
$H_{160}$                                      &  25.7$\pm$0.1       & 25.2$\pm$0.1         & 26.1$\pm$0.1         \\   
\footnotesize[3.6~\micron  ]                   &  25.2$\pm$0.3       & 23.4$\pm$0.1         & 23.1$\pm$0.1         \\ 
\footnotesize[4.5~\micron  ]                   &  24.4$\pm$0.3       & 22.9$\pm$0.1         & 22.8$\pm$0.1         \\ 
$z_{\rm phot,\hst\ only}$$^{\rm a}$            & 9.6$^{+0.7}_{-7.4}$ & 10.0$^{+0.6}_{-7.5}$ & 9.9$^{+0.7}_{-1.0}$  \\ 
$z_{\rm phot,\hst +\spitzer}$\hspace{0.2cm}    & 9.9$^{+0.6}_{-0.6}$ & 2.7$^{+0.1}_{-0.1}$  & 3.6$^{+0.2}_{-0.2}$  \\ [0.2 cm]
\hline
\end{tabular*}
\begin{tablenotes}
\footnotesize
\item {\bf Notes:} $^{a}$Photometric redshifts found using BPZ. 
The two $z\sim3$ interlopers from PLCKG138-10 and RXC0018+16 were initially identified as $z\sim10$ candidates prior to including the \spitzer\ data, whereas the candidate in SPT0615-57 remained at $z\sim10$. 
\end{tablenotes}
\vspace{0.2cm}
\end{table}
\egroup

\begin{figure*}
\label{fig:cluster}
\vspace{0.2cm}
\mbox{\centerline{ \includegraphics[scale=0.90,trim=0pt 0pt 0pt 0pt,clip]{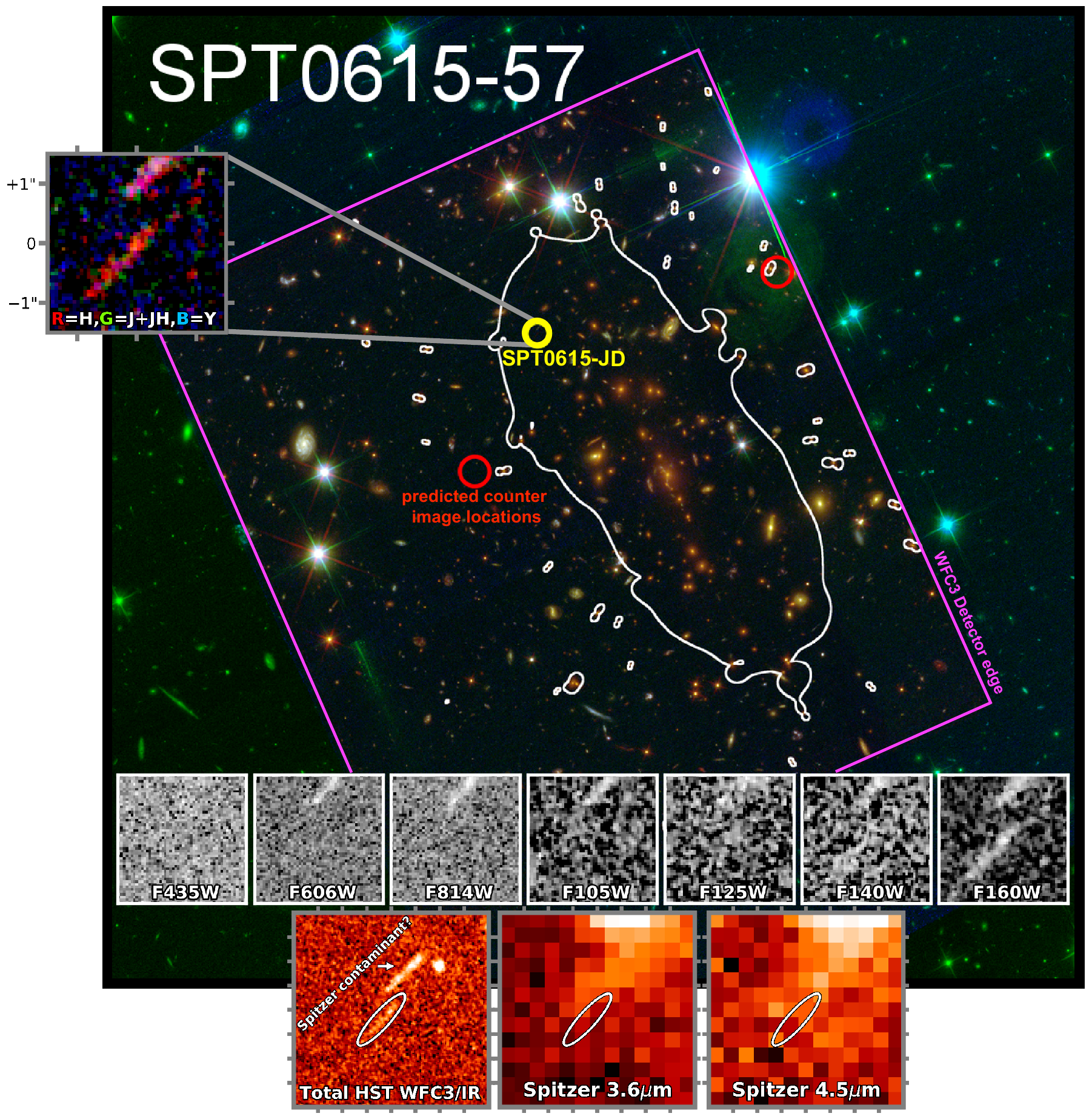}}}
\caption{
A 3\farcm 25$\times$3\farcm 25 color image of the \hst\ RELICS cluster field
SPT0615-57. The yellow circle marks the location of the $z\sim10$ candidate SPT0615-JD.
The white lines and red circles show the lens-model $z=10$ critical curves and predicted locations
of the yet undetected counterimages. The violet lines mark the edge where WFC3/IR data is available
The expanded inset is a 3\arcsec x 3\arcsec\ WFC3/IR RGB color image image with the 
R channel as F160W, G as the sum of F125W and F140W, and B as F105W. 
The bottom row of insets are the ACS images followed by WFC3/IR images, all 3\arcsec x3\arcsec\
and 60~mas resolution.
The candidate is missing in the bands blueward of F140W, indicating a strong spectral break. 
Bottom row: larger 8\arcsec x8\arcsec\ cutouts with ellipses marking the position of SPT0615-JD. 
Bottom left, right, and middle: A weighted stack of all four WFC3/IR bands centered on SPT0615-JD, 
and the \emph{Spitzer} 3.6~\micron\ and 4.5~\micron\ images. 
The \emph{Spitzer} flux from the nearby bright $z\sim3$ galaxy crowds the $z\sim10$ candidate, which
appears otherwise faint. 
}
\end{figure*}

Table~\ref{tab:candidates} shows the three $z>9$ candidates found in RELICS after fitting photometric redshifts
to only \hst\ data. We then vetted these candidates by checking \spitzer\ data from the
S-RELICS programs (PI Brada{\v c}; PI Soifer). The IRAC 
channel 1 and 2 bands (3.6~\micron\ and 4.5~\micron\ respectively, with 
$\approx$5-hour depth per band), correspond to
rest-frame optical flux at $z\sim9-10$ and are invaluable for distinguishing between
intrinsically bluer $z\sim10$ star-forming galaxies and intrinsically redder $z\sim$3 interloper galaxies. 
The \spitzer\ fluxes were extracted using \code{T-PHOT} \citep{Merlin16} which uses the higher-resolution
\hst\ imaging as a prior to extract photometry from the lower resolution \spitzer\ images. First, we produce PSF 
convolution kernels based on all available \hst\ images. We manually sharpen the PSF to minimize
residuals between the convolved image and the \spitzer\ images. Then, we run \code{T-PHOT} on the entire
cluster field to extract the \spitzer\ photometry.

After obtaining the \spitzer\ photometry and re-running the photometric redshifts, 
we rule out two candidates as low-$z$ interlopers, leaving one $z\sim10$ candidate. 
Fig.~\ref{fig:cluster} shows image cutouts of this candidate, hereafter named 
SPT0615-JD (``JD" for \hst\ F125W $J$-band dropout),
in each of the available \hst\ and \spitzer\ bands, as well as a WFC3/IR color composite. 
SPT0615-JD has an AB magnitude of $25.7\pm0.1$ in F160W detected with S/N$\sim$11
(the F160W exposures were in two epochs 44 days apart and each detected the source 
with S/N$\sim$5). The extended arc shape is consistent with
the direction of lensing shear expected from the cluster (see \S \ref{sec:lens}).
The bands blueward of F140W are undetected with S/N$\lesssim$2, 
and F140W is just undetected (S/N=2.9). Importantly, we emphasize 
that observed-frame size of SPT0516-JD is rather large ($\sim$2.5\arcsec\ long), 
and can easily be spatially resolved by \jwst\ (see~\S \ref{sec:compare}). 

We note that these image cutouts reveal an important caveat to the \spitzer\ fluxes of SPT0615-JD.
\code{T-PHOT} reports a maximum covariance between SPT0615-JD
and all other sources fit simultaneously (max CV ratio) of $\sim1.4$ for both the 
3.6~\micron\ and 4.5~\micron\ images. This implies a covariance between the \spitzer\ photometry of 
SPT0615-JD and a nearby source. The 8\arcsec\ x8\arcsec\ \hst\ image in 
Fig.~\ref{fig:cluster} shows an IR-bright nearby 
$z\sim3$ galaxy. We conclude that the \spitzer\ fluxes for SPT0615-JD are biased by this source 
and are likely over-estimated. Even so, the fluxes in each \spitzer\ band are already 
several magnitudes fainter than typical low-$z$ interlopers. This is critical because while all 
$z \sim10$ solutions could have lower \emph{Spitzer} fluxes, the $z\sim2$ solution 
\emph{requires} them to be high, especially at 4.5~\micron. As we will discuss in \S \ref{sec:SED}, 
the inflated \spitzer\ fluxes also increase the inferred $z\sim$10 UV dust attenuation, which should 
be considered an upper limit. Upcoming deeper \spitzer\ imaging (PI~Brada{\v c}) of this 
cluster will improve constraints on the flux and derived properties of this candidate. 

\section{Lens Models}\label{sec:lens}
We identify three sets of multiply-imaged galaxies: two with spectroscopic redshifts
($z=1.358$ and $z=4.013$) and one whose redshift is free to vary in the modeling. 
(Paterno-Mahler et al., in preparation).  Based on these, we 
produce three lens models using \code{Lenstool} \citep{Jullo07}, \code{GLAFIC} \citep{Oguri10}, 
and the \cite{Zitrin15b} Light Traces Mass (LTM) method.  Based on these models, 
we estimate the magnification of SPT0615-JD to be $\mu\sim4-7$.

All three models predict two counterimages at the positions shown in Fig.~\ref{fig:cluster}.  
Our results using \code{GLAFIC} (Kikuchihara et al., in preparation) and \code{Lenstool} 
(Paterno-Mahler et al., in preparation) predict the upper-right counterimage
is $\approx$1 magnitude fainter than the original arc, and therefore below the detection limit. All three models
predict the lower-left counterimage to be of similar magnification and magnitude of the original arc, and
\code{LTM} predicts the counterimages to have the same magnitude as the original arc. 
Given these models, we would have expected to see an image near the lower-left position, but none are
yet detected. We note that the WFC3 limiting depths are $\sim$26 AB mag, and the 
counterimages may be fainter. Conversely, all lens models predict no counterimages if SPT0615-JD is
at $z\sim2$. Deeper imaging of this field is required to properly search for the $z \sim10$ counterimages 
and yield geometric support as in \cite{Coe13}, \cite{Zitrin14}, and \cite{Chan17}.

\begin{figure} 
\label{fig:SEDfit}
\hspace{-0.4cm}
\mbox{ \includegraphics[scale=0.35,trim=1pt 0pt 0pt 0pt,clip]{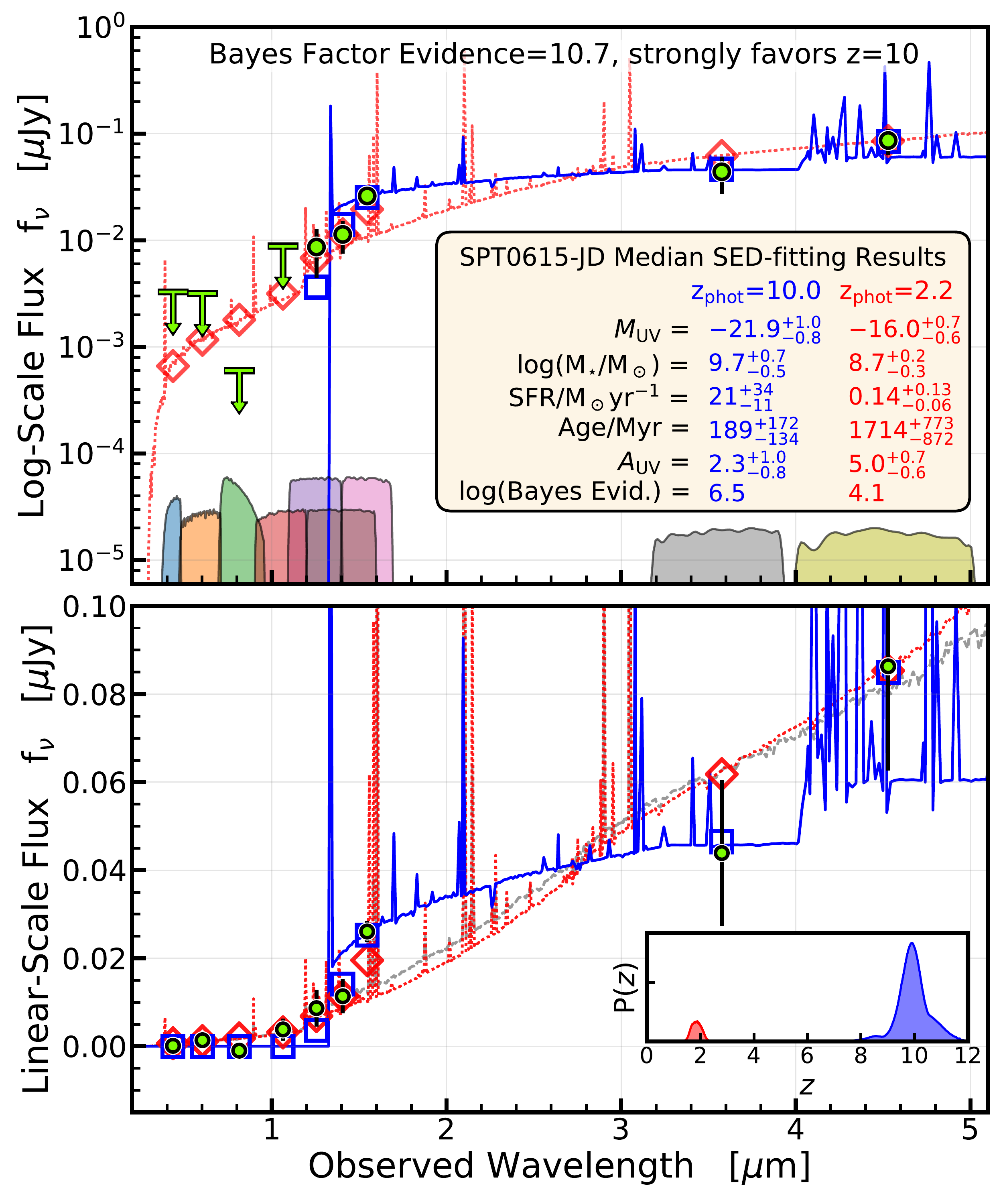}}
\caption{
Best-fit SEDs to RELICS \hst\ and \spitzer\ photometry (green circles) 
of SPT0615-JD. The solid blue line and squares (dotted red line and diamonds) show the 
best-fit SED and model fluxes respectively assuming the $z\sim$ 10 ($z\sim$ 2) solution. The 
top (bottom) plot displays the flux on a logarithmic (linear) scale, and the top plot also shows the \hst\
and \spitzer\ transmission curves for reference. The dashed gray curve in the lower plot shows the 
$z\sim2$ median SED. Fluxes have been corrected for lensing magnification
($\mu=7$) and the \spitzer\ fluxes have not been adjusted despite likely contamination 
(see \S \ref{sec:Data} and Fig.~\ref{fig:cluster}). 
The lower inset figure shows the redshift likelihood P($z$) which strongly favors the $z\sim10$
solution. Though the $z\sim2$ solutions formally have a $\sim10\%$ likelihood, we argue they
are unphysical in the text (\S \ref{sec:SED}).
}
\end{figure}
%

%
\begin{figure*} 
\label{fig:MagVsRedshift}
\centerline{\includegraphics[scale=0.36]{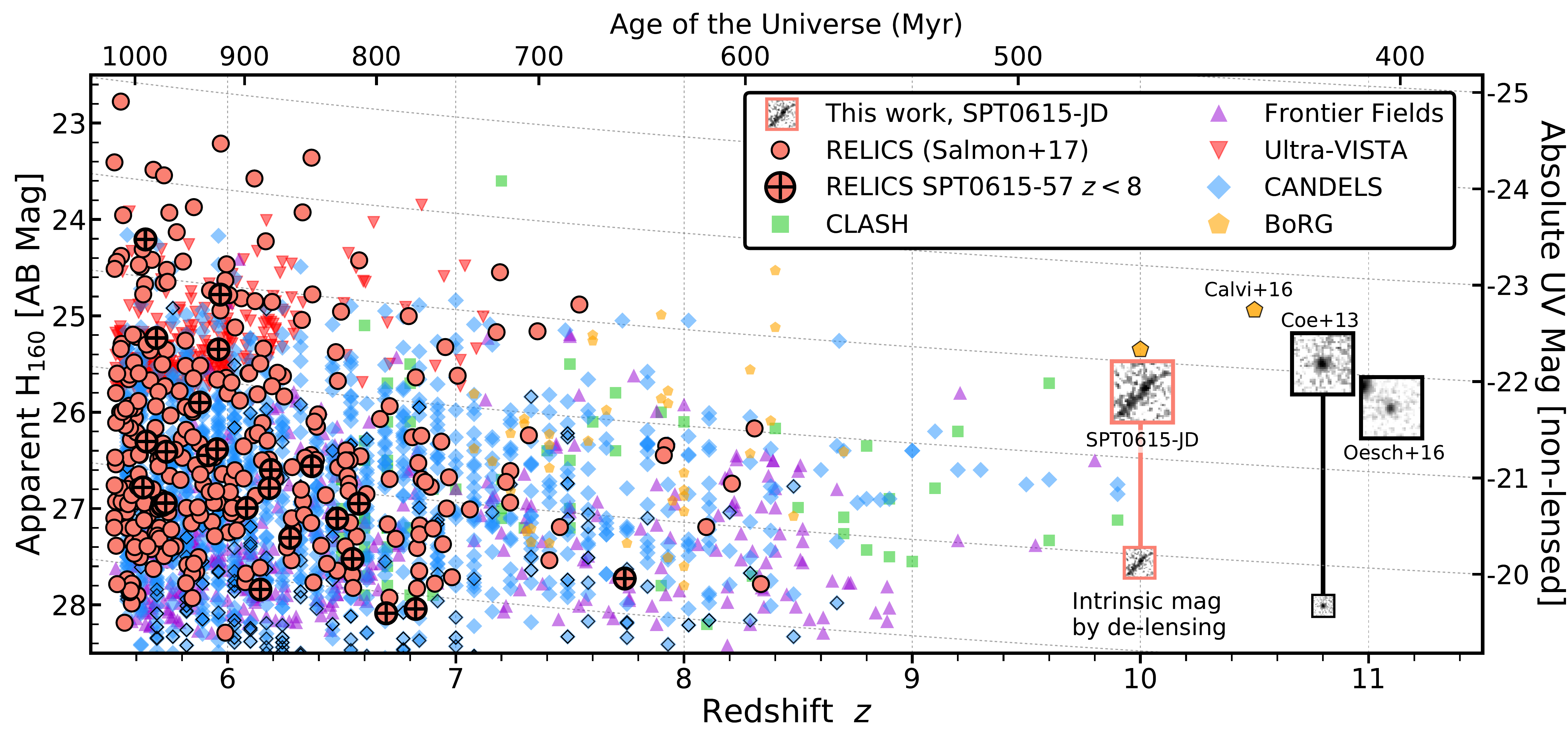}}
\caption{
Observed $H$-band magnitude versus redshift for $z>5.5$ candidates from various surveys. 
The $z<8.5$ candidates from \cite{Salmon17} are shown as salmon-colored circles, 
and the candidates from SPT0615-57 are filled with crosses. 
The green squares are galaxies from CLASH  \citep{Zheng12, Bradley14, Hoag17}, 
purple upwards-triangles from the Frontier Field \citep{Zitrin14, Ishigaki17},
red downwards-triangle from Ultra-VISTA \citep{Bowler17},
blue diamonds from CANDELS \citep{Bouwens14, Bouwens15, Oesch16} 
\citep[outlined diamonds are from the HUDF; see also][]{Finkelstein15}, and
orange pentagons from BoRG/HIPPIES \citep{Bradley12,Schmidt14,Calvi16}.
Gray lines follow the conversion from apparent to absolute UV magnitude to reference 
for un-lensed sources.  2\arcsec x2\arcsec cutout images of two $z\sim11$ candidates 
\citep{Coe13,Oesch16} and the $z\sim10$ candidate from this work
mark their observed magnitudes respectively. 
For the latter two candidates, we also 
show their cutouts scaled in size according to their lens model to show 
an example of their intrinsic size (a full source-plane reconstruction of SPT0615-JD would 
show a smaller axis ratio).
SPT0615-JD has a much larger observed size compared to the other candidates. 
}
\vspace{0.25 cm}
\end{figure*}

\section{SED Fitting}\label{sec:SED}
Thanks to the \spitzer\ data that probes the rest-frame optical and 
near-ultraviolet (UV, ${\sim2900-4500}$~\AA), we can 
infer upper-limits on physical parameters like stellar mass and 
dust attenuation to test if the high and low-redshift solutions are sensible. 
We use a Bayesian SED-fitting procedure originally described by \cite{Papovich01} and 
updated by \cite{Salmon15}. In short, we sample the posterior using a grid of 
SEDs that represent a range of stellar population ages (${10\ {\rm{Myr}}<t_{\rm{age}}<t_{\rm universe}}$, 
logarithmically spaced), 
attenuation (${0<A_{\rm{UV}}<7.4}$),  metallicity (${0.02Z_\Sun<Z<Z_\Sun}$), 
and rising star-formation histories
($\Psi (t) =\Psi_0 \exp(t/\tau_{\rm SFH})$, where the $e$-folding timescale $\tau_{\rm SFH}$ can be 
 0.3, 0.5, 0.7, 1, 3, 5, 7, 10, 30, 50, 70, or 100 Gyr). We use \cite{Bruzual03} stellar population synthesis
models with a \cite{Chabrier03} IMF\footnote{Switching from a Chabrier to a \cite{Salpeter55} IMF would result in 
higher derived stellar mass and star-formation rate (SFR) by 0.25 dex.} and include the effects of nebular emission lines 
following \cite{Salmon15}. We assume the dust-attenuation law derived by \cite{Salmon16} that varies in shape 
from a steep law at low attenuation (similar in shape to the extinction law of the Small Magellanic Cloud)
to a grey law at high attenuation (similar in shape to the starburst curve of \cite{Calzetti00}). 

The results of our SED fitting are summarized in Fig.~\ref{fig:SEDfit}. 
For all SED fitting, we correct
for lensing magnification assuming $\mu=7$, and do not further correct the \spitzer\ fluxes 
despite likely contamination (see \S \ref{sec:Data} and Fig.~\ref{fig:cluster}). 
The fits assuming the $z\sim10$ redshift show a moderately high stellar mass
of $M_\star=10^{9.7^{+0.7}_{-0.5}}\ \rm{M}_\Sun$ and star-formation rate of SFR=21$^{+34}_{-11}$ 
$\rm{M}_\Sun$/yr. 
However, the stellar mass, star-formation rate, age, and UV dust attenuation will be lower
if the rest-frame optical fluxes are over-estimated, as implied by the \spitzer\ contaminant 
shown in Fig.~\ref{fig:cluster}. Therefore, we consider these preliminary estimates to be upper limits.
Nevertheless, the stellar mass and SFR of SPT0615-JD are indicative of a typical star-forming galaxy 
at $z\sim10$ \citep{Oesch14} and would lie on the SFR-\Mstar\ relation at $z\sim6$ \citep{Salmon15}. 

The SED fit assuming $z\sim2$ is quite different. The median, marginalized results, which
account for the full probability density, imply a low stellar mass 
${M_\star=10^{8.7^{+0.2}_{-0.3}}\ \rm{M}_\Sun}$, low star-formation rate SFR= ${0.14^{+0.13}_{-0.06}\ 
\rm{M}_\Sun }$/yr, and evolved stellar population age $t$=1714$^{+773}_{-872}$~Myr, with high uncertainty. 
Similarly, the best-fit SED has the same stellar mass, but a higher SFR (SFR$\sim{3\ \rm{M}_\Sun}$/yr), a slightly 
younger stellar population age ($t$=1585~Myr), a very dusty SED ($A_{\rm UV}=6.1$~mag), and high nebular emission 
([OIII]+H$_\beta$ equivalent width EW=780~\AA, or 1671~\AA to match the $H$-band magnitude). 

This $z\sim2$ SED solution is unphysical for several reasons. Its dust attenuation is dramatically high for its low stellar mass
\citep{Pannella09}, its size is too large and SED too dusty compared to other extreme [OIII] emitting 
galaxies at $z\sim2$ \citep{Malkan17}, and it has too high EW compared 
to [OIII] emitters at $z\sim2$ of similar mass \citep{Maseda13}. Such a rare 
high EW and high dust interloper was spectroscopically ruled out for a similar $z\sim11$ candidate, MACS0646-JD \citep{Pirzkal15}.
Importantly, the $z\sim2$ SED necessitates such strong [OIII] emission to match the observations with appreciable likelihood; 
a dusty SED template alone cannot match both the bright H-band flux and the relatively faint optical flux. 
Finally, unlike the $z\sim10$ solution, the $z\sim2$ solution \emph{requires} the already 
overestimated \spitzer\ fluxes to be high, and it becomes increasingly 
harder to justify a $z\sim2$ SED with lower 3.6~\micron\ and 4.5~\micron\ fluxes. 

We caution the reader that the best-fit SEDs in Fig.~\ref{fig:SEDfit} 
(and best-fit SED solutions in general) are not necessarily representative of the full 
probability density of the posterior \citep{Leja17}. 
A better indicator of the goodness-of-fit than the best-fit $\chi^2$ is the 
unconditional marginal likelihood of the data, or the 
Bayesian evidence \citep[see e.g.,][for definitions]{Salmon16}, which describes 
probability of seeing the data given all parameters. 
The ratio of two Bayesian 
evidences\footnote{The Bayes-factor evidence is typically described as 
$\zeta=2\ln B_{12}$, where $B_{12}$ is the ratio of the Bayesian evidence 
under model assumption~1 to that of model assumption~2. 
Larger positive numbers favor the assumptions in model~1, 
and negative favor the assumptions in model~2 \citep{Kass95}.} 
under different model assumptions, like whether we assume the SED lies at 
$z\sim2$ or $z\sim$10, is called the Bayes Factor and describes the 
relative evidence between two model assumptions. 
We find a Bayes-Factor evidence of 10.7 in favor of the $z\sim10$ solution, 
which is considered ``very strong" evidence \citep{Kass95}. Our interpretation is that there
are more SEDs that fit the data well assuming the $z\sim10$ 
redshift than the select few (and justifiably unphysical) SEDs that fit the data well
assuming the $z\sim2$ redshift. 

\section{Comparison with Other High-Redshift Candidates}\label{sec:compare}
Fig.~\ref{fig:MagVsRedshift} shows the $H$-band magnitude versus redshift
for all high-$z$ ($z>5.5$) candidate galaxies discovered in RELICS \citep{Salmon17}
and many other deep and wide surveys. The lensed, observed-frame size of SPT0615-JD stands 
out as spatially much larger than other $z\sim10$ candidates
(other candidates at these redshifts have similar point-like sizes 
to those found by \citealt{Coe13} and \citealt{Oesch16}, see 
below). The intrinsic (de-lensed) magnitude of SPT0615-JD is similar to
that of the $z\sim11$ candidate MACS0647-JD \citep{Coe13}. 

\begin{figure*} 
\label{fig:Sizes}
\centerline{\mbox{ \includegraphics[scale=0.46,trim=1pt 0pt 0pt 0pt,clip]{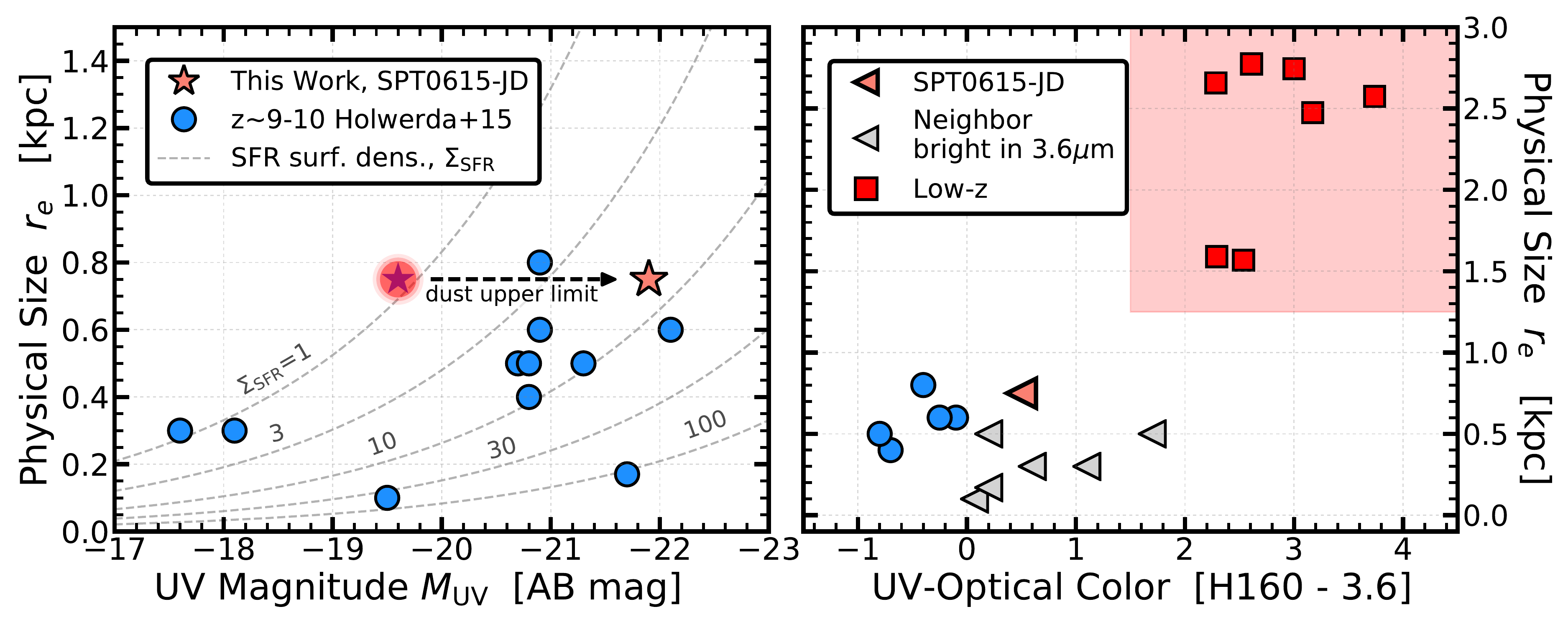}}}
\caption{
The size of SPT0615-JD compared to other known $z>9$ galaxy candidates (blue circles). 
Left: the physical size (effective radius) as a function of absolute UV magnitude. 
The delensed magnitude is shown as the red circle and star,
and the unobscured (corrected for our upper-limit dust extinction) 
delensed magnitude is shown as the salmon-colored star.
Right: the physical size as a function of [F160W$-$3.6~\micron] color. 
Triangles show $z>9$ candidates, including SPT0615-JD (bold triangle), 
that may be contaminated in Spitzer flux by a bright neighbor, 
resulting in redder [F160W$-$3.6~\micron] colors. Significantly larger sizes of 
typical very red $z\sim2$ galaxies (red squares) are shown in the red box. 
}
\vspace{0.25cm}
\end{figure*}
%

An independent way to test high and low-$z$ solutions for SPT0615-JD 
is to calculate its physical size and compare to other known
interlopers. Moreover, the sizes of galaxies can give us great physical 
insight into the initial conditions of early disk evolution \citep{Ferguson04}. 
Broadly, the $z>5$ size evolution at fixed luminosity scales as (1+$z$)$^{-m}$ where 
$m=1-2$ \citep{Shibuya15}. \cite{Holwerda15} demonstrated 
that a combination of UV-to-optical color, sampled by the F160W and 3.6~\micron\ bands, 
and physical size can be used to identify obvious low-$z$ contaminants.
They summarized that the sizes of $z>9$ galaxy candidates have typical 
half-light radii of $r_e<0.8$~kpc. 

To calculate the size of SPT0615-JD, we used our lens models 
to reconstruct its image in the source plane. The LTM lens model finds a 
relatively mild tangential magnification, or shear, of a factor of $\sim3$, leaving 
the full width of the de-lensed source to be about 3--3.5~kpc. 
If we assume the light distribution is uniform, we can take the half light radius to be 
about $\sim$1/4 of the full size and find $r_e\approx0.7-0.8$~kpc. The
statistical error on this size (from the lens model) is only a couple of percent, 
so we are dominated by systematic errors ($\sim$10\%). Curiously, the reconstructed 
source's axis ratio is still about 2:1 in the same direction as the lensing shear, 
which could mean that the shear is underestimated and the size is in fact smaller. 

Fig.~\ref{fig:Sizes} shows that the inferred size of SPT0615-JD is typical 
compared to other high-$z$ candidates. This provides crucial evidence in
support of the $z\sim10$ solution that is independent of the galaxy SED. While 
the uncertainty in the $z\sim10$ UV dust attenuation should be considered as an
upper limit, the candidate is still within the range of $M_{\rm UV}$ and SFR 
surface density of known $z>9$ candidates. 

\section{Conclusions}
We present SPT0615-JD, a promising $z\sim10$ galaxy candidate that appears to be 
stretched into the shape of an arc by 
the effects of strong gravitational lensing. Out of all
combined lensing fields from RELICS, CLASH, and the Frontier Fields,
there is no other galaxy candidate spatially stretched by lensing as 
distant as SPT0615-JD. 
While our three independent lens models predict at least one 
detectable counterimage, we do not see one in the current data. No counterimages 
are expected if the candidate is at lower redshift. 
However, the only $z\sim2$ SED that fits the data well is unphysical based on the required 
combination of its size, mass, dust attenuation, and [OIII]+H$_\beta$~EW. 
In addition, we find very strong Bayesian evidence that the SED-inferred physical properties 
of this candidate are of a $z\sim10$ typical star-forming galaxy. 
Finally, the source-plane size of SPT0615-JD is similar to other $z=9-10$ galaxies, 
while the observed-frame image offers unprecedented spatial resolution. This 
galaxy candidate offers the unique opportunity for resolving
stellar populations deep in the epoch of reionization, especially with the higher 
resolution of \jwst. 

\section*{Acknowledgements} 
This paper uses observations from NASA/ESA \hst.  
STScI is operated by AURA under NASA contract NAS 5- 26555. ACS 
under NASA contract NAS 5-32864, and \emph{Spitzer}
by JPL. These observations are associated with program GO-14096 and
archival data with GO-9270, GO-12166, GO-12477, GO-12253.  Some data
were obtained from MAST. This work was performed under the auspices
of the U.S. Department of Energy by LLNL under contract DE-AC52-07NA27344. 
F.A.-S. acknowledges support from Chandra grant G03-14131X.

\bibliographystyle{apj} \bibliography{ms}

\begin{thebibliography}{}
\expandafter\ifx\csname natexlab\endcsname\relax\def\natexlab#1{#1}\fi

\bibitem[{{Bertin} \& {Arnouts}(1996)}]{Bertin96}
{Bertin}, E., \& {Arnouts}, S. 1996, \aaps, 117, 393

\bibitem[{{Bouwens} {et~al.}(2014){Bouwens}, {Bradley}, {Zitrin}, {Coe},
  {Franx}, {Zheng}, {Smit}, {Host}, {Postman}, {Moustakas}, {Labb{\'e}},
  {Carrasco}, {Molino}, {Donahue}, {Kelson}, {Meneghetti}, {Ben{\'{\i}}tez},
  {Lemze}, {Umetsu}, {Broadhurst}, {Moustakas}, {Rosati}, {Jouvel},
  {Bartelmann}, {Ford}, {Graves}, {Grillo}, {Infante}, {Jimenez-Teja}, {Lahav},
  {Maoz}, {Medezinski}, {Melchior}, {Merten}, {Nonino}, {Ogaz}, \&
  {Seitz}}]{Bouwens14}
{Bouwens}, R.~J., {Bradley}, L., {Zitrin}, A., {et~al.} 2014, \apj, 795, 126

\bibitem[{{Bouwens} {et~al.}(2015){Bouwens}, {Illingworth}, {Oesch}, {Trenti},
  {Labb{\'e}}, {Bradley}, {Carollo}, {van Dokkum}, {Gonzalez}, {Holwerda},
  {Franx}, {Spitler}, {Smit}, \& {Magee}}]{Bouwens15}
{Bouwens}, R.~J., {Illingworth}, G.~D., {Oesch}, P.~A., {et~al.} 2015, \apj,
  803, 34

\bibitem[{{Bowler} {et~al.}(2017){Bowler}, {Dunlop}, {McLure}, \&
  {McLeod}}]{Bowler17}
{Bowler}, R.~A.~A., {Dunlop}, J.~S., {McLure}, R.~J., \& {McLeod}, D.~J. 2017,
  \mnras, 466, 3612

\bibitem[{{Bradley} {et~al.}(2012){Bradley}, {Trenti}, {Oesch}, {Stiavelli},
  {Treu}, {Bouwens}, {Shull}, {Holwerda}, \& {Pirzkal}}]{Bradley12}
{Bradley}, L.~D., {Trenti}, M., {Oesch}, P.~A., {et~al.} 2012, \apj, 760, 108

\bibitem[{{Bradley} {et~al.}(2014){Bradley}, {Zitrin}, {Coe}, {Bouwens},
  {Postman}, {Balestra}, {Grillo}, {Monna}, {Rosati}, {Seitz}, {Host}, {Lemze},
  {Moustakas}, {Moustakas}, {Shu}, {Zheng}, {Broadhurst}, {Carrasco}, {Jouvel},
  {Koekemoer}, {Medezinski}, {Meneghetti}, {Nonino}, {Smit}, {Umetsu},
  {Bartelmann}, {Ben{\'{\i}}tez}, {Donahue}, {Ford}, {Infante}, {Jimenez-Teja},
  {Kelson}, {Lahav}, {Maoz}, {Melchior}, {Merten}, \& {Molino}}]{Bradley14}
{Bradley}, L.~D., {Zitrin}, A., {Coe}, D., {et~al.} 2014, \apj, 792, 76

\bibitem[{{Bruzual} \& {Charlot}(2003)}]{Bruzual03}
{Bruzual}, G., \& {Charlot}, S. 2003, \mnras, 344, 1000

\bibitem[{{Calvi} {et~al.}(2016){Calvi}, {Trenti}, {Stiavelli}, {Oesch},
  {Bradley}, {Schmidt}, {Coe}, {Brammer}, {Bernard}, {Bouwens}, {Carrasco},
  {Carollo}, {Holwerda}, {MacKenty}, {Mason}, {Shull}, \& {Treu}}]{Calvi16}
{Calvi}, V., {Trenti}, M., {Stiavelli}, M., {et~al.} 2016, \apj, 817, 120

\bibitem[{{Calzetti} {et~al.}(2000){Calzetti}, {Armus}, {Bohlin}, {Kinney},
  {Koornneef}, \& {Storchi-Bergmann}}]{Calzetti00}
{Calzetti}, D., {Armus}, L., {Bohlin}, R.~C., {et~al.} 2000, \apj, 533, 682

\bibitem[{{Chabrier}(2003)}]{Chabrier03}
{Chabrier}, G. 2003, \pasp, 115, 763

\bibitem[{{Chan} {et~al.}(2017){Chan}, {Broadhurst}, {Lim}, {Diego}, {Zitrin},
  {Coe}, \& {Ford}}]{Chan17}
{Chan}, B.~M.~Y., {Broadhurst}, T., {Lim}, J., {et~al.} 2017, \apj, 835, 44

\bibitem[{{Coe} {et~al.}(2013){Coe}, {Zitrin}, {Carrasco}, {Shu}, {Zheng},
  {Postman}, {Bradley}, {Koekemoer}, {Bouwens}, {Broadhurst}, {Monna}, {Host},
  {Moustakas}, {Ford}, {Moustakas}, {van der Wel}, {Donahue}, {Rodney},
  {Ben{\'{\i}}tez}, {Jouvel}, {Seitz}, {Kelson}, \& {Rosati}}]{Coe13}
{Coe}, D., {Zitrin}, A., {Carrasco}, M., {et~al.} 2013, \apj, 762, 32

\bibitem[{{Ferguson} {et~al.}(2004){Ferguson}, {Dickinson}, {Giavalisco},
  {Kretchmer}, {Ravindranath}, {Idzi}, {Taylor}, {Conselice}, {Fall},
  {Gardner}, {Livio}, {Madau}, {Moustakas}, {Papovich}, {Somerville},
  {Spinrad}, \& {Stern}}]{Ferguson04}
{Ferguson}, H.~C., {Dickinson}, M., {Giavalisco}, M., {et~al.} 2004, \apjl,
  600, L107

\bibitem[{{Finkelstein}(2016)}]{Finkelstein16}
{Finkelstein}, S.~L. 2016, PASA, 33, e037

\bibitem[{{Finkelstein} {et~al.}(2015){Finkelstein}, {Ryan}, {Papovich},
  {Dickinson}, {Song}, {Somerville}, {Ferguson}, {Salmon}, {Giavalisco},
  {Koekemoer}, {Ashby}, {Behroozi}, {Castellano}, {Dunlop}, {Faber}, {Fazio},
  {Fontana}, {Grogin}, {Hathi}, {Jaacks}, {Kocevski}, {Livermore}, {McLure},
  {Merlin}, {Mobasher}, {Newman}, {Rafelski}, {Tilvi}, \&
  {Willner}}]{Finkelstein15}
{Finkelstein}, S.~L., {Ryan}, Jr., R.~E., {Papovich}, C., {et~al.} 2015, \apj,
  810, 71

\bibitem[{{Hoag} {et~al.}(2017){Hoag}, {Brada{\v c}}, {Trenti}, {Treu},
  {Schmidt}, {Huang}, {Lemaux}, {He}, {Bernard}, {Abramson}, {Mason},
  {Morishita}, {Pentericci}, \& {Schrabback}}]{Hoag17}
{Hoag}, A., {Brada{\v c}}, M., {Trenti}, M., {et~al.} 2017, Nature Astronomy,
  1, 0091

\bibitem[{{Holwerda} {et~al.}(2015){Holwerda}, {Bouwens}, {Oesch}, {Smit},
  {Illingworth}, \& {Labbe}}]{Holwerda15}
{Holwerda}, B.~W., {Bouwens}, R., {Oesch}, P., {et~al.} 2015, \apj, 808, 6

\bibitem[{{Infante} {et~al.}(2015){Infante}, {Zheng}, {Laporte}, {Troncoso
  Iribarren}, {Molino}, {Diego}, {Bauer}, {Zitrin}, {Moustakas}, {Huang},
  {Shu}, {Bina}, {Brammer}, {Broadhurst}, {Ford}, {Garc{\'{\i}}a}, \&
  {Kim}}]{Infante15}
{Infante}, L., {Zheng}, W., {Laporte}, N., {et~al.} 2015, \apj, 815, 18

\bibitem[{{Ishigaki} {et~al.}(2017){Ishigaki}, {Kawamata}, {Ouchi}, {Oguri}, \&
  {Shimasaku}}]{Ishigaki17}
{Ishigaki}, M., {Kawamata}, R., {Ouchi}, M., {Oguri}, M., \& {Shimasaku}, K.
  2017, ArXiv, 1702.04867

\bibitem[{{Jullo} {et~al.}(2007){Jullo}, {Kneib}, {Limousin},
  {El{\'{\i}}asd{\'o}ttir}, {Marshall}, \& {Verdugo}}]{Jullo07}
{Jullo}, E., {Kneib}, J.-P., {Limousin}, M., {et~al.} 2007, New Journal of
  Physics, 9, 447

\bibitem[{Kass \& Raftery(1995)}]{Kass95}
Kass, R.~E., \& Raftery, A.~E. 1995, Journal of the American Statistical
  Association, 90, 773

\bibitem[{{Laporte} {et~al.}(2017){Laporte}, {Ellis}, {Boone}, {Bauer},
  {Qu{\'e}nard}, {Roberts-Borsani}, {Pell{\'o}}, {P{\'e}rez-Fournon}, \&
  {Streblyanska}}]{Laporte17}
{Laporte}, N., {Ellis}, R.~S., {Boone}, F., {et~al.} 2017, \apjl, 837, L21

\bibitem[{{Leja} {et~al.}(2017){Leja}, {Johnson}, {Conroy}, {van Dokkum}, \&
  {Byler}}]{Leja17}
{Leja}, J., {Johnson}, B.~D., {Conroy}, C., {van Dokkum}, P.~G., \& {Byler}, N.
  2017, \apj, 837, 170

\bibitem[{{Mainali} {et~al.}(2017){Mainali}, {Kollmeier}, {Stark}, {Simcoe},
  {Walth}, {Newman}, \& {Miller}}]{Mainali17}
{Mainali}, R., {Kollmeier}, J.~A., {Stark}, D.~P., {et~al.} 2017, \apjl, 836,
  L14

\bibitem[{{Malkan} {et~al.}(2017){Malkan}, {Cohen}, {Maruyama}, {Kashikawa},
  {Ly}, {Ishikawa}, {Shimasaku}, {Hayashi}, \& {Motohara}}]{Malkan17}
{Malkan}, M.~A., {Cohen}, D.~P., {Maruyama}, M., {et~al.} 2017, \apj, 850, 5

\bibitem[{{Maseda} {et~al.}(2013){Maseda}, {van der Wel}, {da Cunha}, {Rix},
  {Pacifici}, {Momcheva}, {Brammer}, {Franx}, {van Dokkum}, {Bell},
  {Fumagalli}, {Grogin}, {Kocevski}, {Koekemoer}, {Lundgren}, {Marchesini},
  {Nelson}, {Patel}, {Skelton}, {Straughn}, {Drumpf}, {Weiner}, {Whitaker}, \&
  {Wuyts}}]{Maseda13}
{Maseda}, M.~V., {van der Wel}, A., {da Cunha}, E., {et~al.} 2013, \apjl, 778,
  L22

\bibitem[{{Merlin} {et~al.}(2016){Merlin}, {Bourne}, {Castellano}, {Ferguson},
  {Wang}, {Derriere}, {Dunlop}, {Elbaz}, \& {Fontana}}]{Merlin16}
{Merlin}, E., {Bourne}, N., {Castellano}, M., {et~al.} 2016, \aap, 595, A97

\bibitem[{{Oesch} {et~al.}(2017){Oesch}, {Bouwens}, {Illingworth}, {Labbe}, \&
  {Stefanon}}]{Oesch17}
{Oesch}, P.~A., {Bouwens}, R.~J., {Illingworth}, G.~D., {Labbe}, I., \&
  {Stefanon}, M. 2017, ArXiv, 1710.11131

\bibitem[{{Oesch} {et~al.}(2014){Oesch}, {Bouwens}, {Illingworth}, {Labb{\'e}},
  {Smit}, {Franx}, {van Dokkum}, {Momcheva}, {Ashby}, {Fazio}, {Huang},
  {Willner}, {Gonzalez}, {Magee}, {Trenti}, {Brammer}, {Skelton}, \&
  {Spitler}}]{Oesch14}
{Oesch}, P.~A., {Bouwens}, R.~J., {Illingworth}, G.~D., {et~al.} 2014, \apj,
  786, 108

\bibitem[{{Oesch} {et~al.}(2016){Oesch}, {Brammer}, {van Dokkum},
  {Illingworth}, {Bouwens}, {Labb{\'e}}, {Franx}, {Momcheva}, {Ashby}, {Fazio},
  {Gonzalez}, {Holden}, {Magee}, {Skelton}, {Smit}, {Spitler}, {Trenti}, \&
  {Willner}}]{Oesch16}
{Oesch}, P.~A., {Brammer}, G., {van Dokkum}, P.~G., {et~al.} 2016, \apj, 819,
  129

\bibitem[{{Oguri}(2010)}]{Oguri10}
{Oguri}, M. 2010, \pasj, 62, 1017

\bibitem[{{Pannella} {et~al.}(2009){Pannella}, {Carilli}, {Daddi}, {McCracken},
  {Owen}, {Renzini}, {Strazzullo}, {Civano}, {Koekemoer}, {Schinnerer},
  {Scoville}, {Smol{\v c}i{\'c}}, {Taniguchi}, {Aussel}, {Kneib}, {Ilbert},
  {Mellier}, {Salvato}, {Thompson}, \& {Willott}}]{Pannella09}
{Pannella}, M., {Carilli}, C.~L., {Daddi}, E., {et~al.} 2009, \apjl, 698, L116

\bibitem[{{Papovich} {et~al.}(2001){Papovich}, {Dickinson}, \&
  {Ferguson}}]{Papovich01}
{Papovich}, C., {Dickinson}, M., \& {Ferguson}, H.~C. 2001, \apj, 559, 620

\bibitem[{{Pirzkal} {et~al.}(2015){Pirzkal}, {Coe}, {Frye}, {Brammer},
  {Moustakas}, {Rothberg}, {Broadhurst}, {Bouwens}, {Bradley}, {van der Wel},
  {Kelson}, {Donahue}, {Zitrin}, {Moustakas}, \& {Barker}}]{Pirzkal15}
{Pirzkal}, N., {Coe}, D., {Frye}, B.~L., {et~al.} 2015, \apj, 804, 11

\bibitem[{{Planck Collaboration} {et~al.}(2011){Planck Collaboration},
  {Aghanim}, {Arnaud}, {Ashdown}, {Atrio-Barandela}, {Aumont}, {Baccigalupi},
  {Balbi}, {Banday}, {Barreiro}, {Bartlett}, {Battaner}, {Benabed},
  {Beno{\^i}t}, {Bernard}, {Bersanelli}, {Bhatia}, {B{\"o}hringer}, {Bonaldi},
  {Bond}, {Borgani}, {Borrill}, {Bouchet}, {Brown}, {Burigana}, {Cabella},
  {Cantalupo}, {Cappellini}, {Carvalho}, {Catalano}, {Cay{\'o}n}, {Chiang},
  {Chiang}, {Chon}, {Christensen}, {Churazov}, {Clements}, {Colafrancesco},
  {Colombi}, {Crill}, {Cuttaia}, {da Silva}, {Dahle}, {Danese}, {'Arcangelo},
  {Davis}, {de Bernardis}, {de Gasperis}, {de Zotti}, {Delabrouille},
  {Delouis}, {D{\'e}mocl{\`e}s}, {D{\'e}sert}, {Dickinson}, {Diego}, {Dole},
  {Donzelli}, {Dor{\'e}}, {Douspis}, {Dupac}, {Efstathiou}, {En{\ss}lin},
  {Eriksen}, {Finelli}, {Flores-Cacho}, {Forni}, {Fosalba}, {Frailis},
  {Franceschi}, {Fromenteau}, {Galeotta}, {Ganga}, {G{\'e}nova-Santos},
  {Giard}, {Gonz{\'a}lez-Nuevo}, {Gonz{\'a}lez-Riestra}, {G{\'o}rski},
  {Gregorio}, {Gruppuso}, {Hansen}, {Harrison}, {Hein{\"a}m{\"a}ki},
  {Hern{\'a}ndez-Monteagudo}, {Hildebrandt}, {Hivon}, {Hobson}, {Hurier},
  {Jaffe}, {Jones}, {Juvela}, {Keih{\"a}nen}, {Keskitalo}, {Kisner}, {Kneissl},
  {Kurki-Suonio}, {Lagache}, {L{\"a}hteenm{\"a}ki}, {Lamarre}, {Lasenby},
  {Lawrence}, {Le Jeune}, {Leach}, {Leonardi}, {Leroy}, {Liddle}, {Lilje},
  {L{\'o}pez-Caniego}, {Luzzi}, {Mac{\'{\i}}as-P{\'e}rez}, {Maino},
  {Mandolesi}, {Marleau}, {Mart{\'{\i}}nez-Gonz{\'a}lez}, {Masi}, {Matarrese},
  {Mazzotta}, {Meinhold}, {Melchiorri}, {Melin}, {Mendes}, {Mennella},
  {Miville-Desch{\^e}nes}, {Moneti}, {Montier}, {Morgante}, {Mortlock},
  {Munshi}, {Naselsky}, {Natoli}, {Nevalainen}, {N{\o}rgaard-Nielsen},
  {Noviello}, {Novikov}, {Novikov}, {O'Dwyer}, {Osborne}, {Paladini}, {Pasian},
  {Patanchon}, {Pearson}, {Perdereau}, {Perotto}, {Perrotta}, {Piacentini},
  {Pierpaoli}, {Piffaretti}, {Platania}, {Pointecouteau}, {Polenta},
  {Ponthieu}, {Popa}, {Poutanen}, {Pratt}, {Pr{\'e}zeau}, {Prunet}, {Puget},
  {Rachen}, {Rebolo}, {Reinecke}, {Renault}, {Ricciardi}, {Riller},
  {Ristorcelli}, {Rocha}, {Rubi{\~n}o-Mart{\'{\i}}n}, {Saar}, {Sandri},
  {Savini}, {Schaefer}, {Scott}, {Smoot}, {Starck}, {Sutton}, {Sygnet},
  {Tauber}, {Terenzi}, {Toffolatti}, {Tomasi}, {Tristram}, {T{\"u}rler},
  {Valenziano}, {Vielva}, {Villa}, {Vittorio}, {Wade}, {Wandelt}, {Weller},
  {White}, {White}, {Yvon}, {Zacchei}, \& {Zonca}}]{Planck11}
{Planck Collaboration}, {Aghanim}, N., {Arnaud}, M., {et~al.} 2011, \aap, 536,
  A26

\bibitem[{{Rigby} {et~al.}(2015){Rigby}, {Bayliss}, {Gladders}, {Sharon},
  {Wuyts}, {Dahle}, {Johnson}, \& {Pe{\~n}a-Guerrero}}]{Rigby15}
{Rigby}, J.~R., {Bayliss}, M.~B., {Gladders}, M.~D., {et~al.} 2015, \apjl, 814,
  L6

\bibitem[{{Salmon} {et~al.}(2015){Salmon}, {Papovich}, {Finkelstein}, {Tilvi},
  {Finlator}, {Behroozi}, {Dahlen}, {Dav{\'e}}, {Dekel}, {Dickinson},
  {Ferguson}, {Giavalisco}, {Long}, {Lu}, {Mobasher}, {Reddy}, {Somerville}, \&
  {Wechsler}}]{Salmon15}
{Salmon}, B., {Papovich}, C., {Finkelstein}, S.~L., {et~al.} 2015, \apj, 799,
  183

\bibitem[{{Salmon} {et~al.}(2016){Salmon}, {Papovich}, {Long}, {Willner},
  {Finkelstein}, {Ferguson}, {Dickinson}, {Duncan}, {Faber}, {Hathi},
  {Koekemoer}, {Kurczynski}, {Newman}, {Pacifici}, {P{\'e}rez-Gonz{\'a}lez}, \&
  {Pforr}}]{Salmon16}
{Salmon}, B., {Papovich}, C., {Long}, J., {et~al.} 2016, \apj, 827, 20

\bibitem[{{Salmon} {et~al.}(2017){Salmon}, {Coe}, {Bradley}, {Bouwens},
  {Bradac}, {Huang}, {Oesch}, {Stark}, {Sharon}, {Trenti}, {Avila}, {Ogaz},
  {Andrade-Santos}, {Carrasco}, {Cerny}, {Dawson}, {Frye}, {Hoag}, {Johnson},
  {Jones}, {Lam}, {Lovisari}, {Mainali}, {Past}, {Paterno-Mahler}, {Peterson},
  {Reiss}, {Rodney}, {Ryan}, {Sendra-Server}, {Strolger}, {Umetsu}, {Vulcani},
  \& {Zitrin}}]{Salmon17}
{Salmon}, B., {Coe}, D., {Bradley}, L., {et~al.} 2017, ArXiv, 1710.08930

\bibitem[{{Salpeter}(1955)}]{Salpeter55}
{Salpeter}, E.~E. 1955, \apj, 121, 161

\bibitem[{{Schmidt} {et~al.}(2014){Schmidt}, {Treu}, {Trenti}, {Bradley},
  {Kelly}, {Oesch}, {Holwerda}, {Shull}, \& {Stiavelli}}]{Schmidt14}
{Schmidt}, K.~B., {Treu}, T., {Trenti}, M., {et~al.} 2014, \apj, 786, 57

\bibitem[{{Shibuya} {et~al.}(2015){Shibuya}, {Ouchi}, \&
  {Harikane}}]{Shibuya15}
{Shibuya}, T., {Ouchi}, M., \& {Harikane}, Y. 2015, \apjs, 219, 15

\bibitem[{{Smit} {et~al.}(2017){Smit}, {Swinbank}, {Massey}, {Richard},
  {Smail}, \& {Kneib}}]{Smit17a}
{Smit}, R., {Swinbank}, A.~M., {Massey}, R., {et~al.} 2017, \mnras, 467, 3306

\bibitem[{{Stark}(2016)}]{Stark16}
{Stark}, D.~P. 2016, \araa, 54, 761

\bibitem[{{Stark} {et~al.}(2014){Stark}, {Richard}, {Siana}, {Charlot},
  {Freeman}, {Gutkin}, {Wofford}, {Robertson}, {Amanullah}, {Watson}, \&
  {Milvang-Jensen}}]{Stark14}
{Stark}, D.~P., {Richard}, J., {Siana}, B., {et~al.} 2014, \mnras, 445, 3200

\bibitem[{{Stark} {et~al.}(2015){Stark}, {Walth}, {Charlot}, {Cl{\'e}ment},
  {Feltre}, {Gutkin}, {Richard}, {Mainali}, {Robertson}, {Siana}, {Tang}, \&
  {Schenker}}]{Stark15b}
{Stark}, D.~P., {Walth}, G., {Charlot}, S., {et~al.} 2015, \mnras, 454, 1393

\bibitem[{{Williamson} {et~al.}(2011){Williamson}, {Benson}, {High},
  {Vanderlinde}, {Ade}, {Aird}, {Andersson}, {Armstrong}, {Ashby}, {Bautz},
  {Bazin}, {Bertin}, {Bleem}, {Bonamente}, {Brodwin}, {Carlstrom}, {Chang},
  {Chapman}, {Clocchiatti}, {Crawford}, {Crites}, {de Haan}, {Desai}, {Dobbs},
  {Dudley}, {Fazio}, {Foley}, {Forman}, {Garmire}, {George}, {Gladders},
  {Gonzalez}, {Halverson}, {Holder}, {Holzapfel}, {Hoover}, {Hrubes}, {Jones},
  {Joy}, {Keisler}, {Knox}, {Lee}, {Leitch}, {Lueker}, {Luong-Van}, {Marrone},
  {McMahon}, {Mehl}, {Meyer}, {Mohr}, {Montroy}, {Murray}, {Padin}, {Plagge},
  {Pryke}, {Reichardt}, {Rest}, {Ruel}, {Ruhl}, {Saliwanchik}, {Saro},
  {Schaffer}, {Shaw}, {Shirokoff}, {Song}, {Spieler}, {Stalder}, {Stanford},
  {Staniszewski}, {Stark}, {Story}, {Stubbs}, {Vieira}, {Vikhlinin}, \&
  {Zenteno}}]{Williamson11}
{Williamson}, R., {Benson}, B.~A., {High}, F.~W., {et~al.} 2011, \apj, 738, 139

\bibitem[{{Zheng} {et~al.}(2012){Zheng}, {Postman}, {Zitrin}, {Moustakas},
  {Shu}, {Jouvel}, {H{\o}st}, {Molino}, {Bradley}, {Coe}, {Moustakas},
  {Carrasco}, {Ford}, {Ben{\'{\i}}tez}, {Lauer}, {Seitz}, {Bouwens},
  {Koekemoer}, {Medezinski}, {Bartelmann}, {Broadhurst}, {Donahue}, {Grillo},
  {Infante}, {Jha}, {Kelson}, {Lahav}, {Lemze}, {Melchior}, {Meneghetti},
  {Merten}, {Nonino}, {Ogaz}, {Rosati}, {Umetsu}, \& {van der Wel}}]{Zheng12}
{Zheng}, W., {Postman}, M., {Zitrin}, A., {et~al.} 2012, \nat, 489, 406

\bibitem[{{Zitrin} {et~al.}(2014){Zitrin}, {Zheng}, {Broadhurst}, {Moustakas},
  {Lam}, {Shu}, {Huang}, {Diego}, {Ford}, {Lim}, {Bauer}, {Infante}, {Kelson},
  \& {Molino}}]{Zitrin14}
{Zitrin}, A., {Zheng}, W., {Broadhurst}, T., {et~al.} 2014, \apjl, 793, L12

\bibitem[{{Zitrin} {et~al.}(2015){Zitrin}, {Labb{\'e}}, {Belli}, {Bouwens},
  {Ellis}, {Roberts-Borsani}, {Stark}, {Oesch}, \& {Smit}}]{Zitrin15b}
{Zitrin}, A., {Labb{\'e}}, I., {Belli}, S., {et~al.} 2015, \apjl, 810, L12

\end{thebibliography}
\end{document}